\documentclass[aps,PRL,twocolumn,superscriptaddress,floats]{revtex4}
\usepackage{txfonts}
\usepackage{amssymb}
\usepackage{graphicx}
\usepackage{sidecap}

\begin{document}

\title{Nodal superconducting gap structure in ferropnictide superconductor BaFe$_2$(As$_{0.7}$P$_{0.3}$)$_2$}

\author{Y. Zhang}\author{Z. R. Ye}\author{Q. Q. Ge}\author{F. Chen}\author{Juan Jiang}\author{M. Xu}\author{B. P. Xie}\author{D. L. Feng}\email{dlfeng@fudan.edu.cn}
\affiliation{State Key Laboratory of Surface Physics, Department of
Physics, and Advanced Materials Laboratory, Fudan University,
Shanghai 200433, People's Republic of China}

\maketitle

\textbf{The superconducting gap is a pivotal character for a
superconductor.  While the cuprates and conventional phonon-mediated
superconductors are characterized by distinct $d$-wave and $s$-wave
pairing symmetry with nodal and nodeless gap distribution
respectively, the superconducting gap distributions in iron-based
superconductors are rather diversified.  While nodeless gap
distributions have been directly observed in
Ba$_{1-x}$K$_x$Fe$_2$As$_2$,  BaFe$_{2-x}$Co$_{x}$As$_2$,
K$_x$Fe$_{2-y}$Se$_2$, and FeTe$_{1-x}$Se$_x$
\cite{nodeless1,nodeless2,ZhangKFe2Se2,DingFeSe}, the signatures of
nodal superconducting gap have been reported in LaOFeP, LiFeP,
KFe$_2$As$_2$, BaFe$_2$(As$_{1-x}$P$_x$)$_2$,
BaFe$_{2-x}$Ru$_x$As$_2$ and FeSe \cite{LaFePO,linenode1,
linenode2,linenode3,LiFeP,ShiyanKFeAs,ShiyanBaRu,XueFeSeSTM}. We
here report the angle resolved photoemission spectroscopy (ARPES)
measurements on the superconducting gap structure of
BaFe$_2$(As$_{1-x}$P$_x$)$_2$ in the momentum space, and
particularly, the first direct observation of a circular line node
on the largest hole Fermi surface around the Z point at the
Brillouin zone boundary. Our data rules out the $d$-wave pairing
origin of the nodal gap, and unify both the nodal and nodeless gaps
in iron pnictides under the $s^{\pm}$ pairing symmetry.}

\begin{figure*}[t!]
\includegraphics[width=17cm]{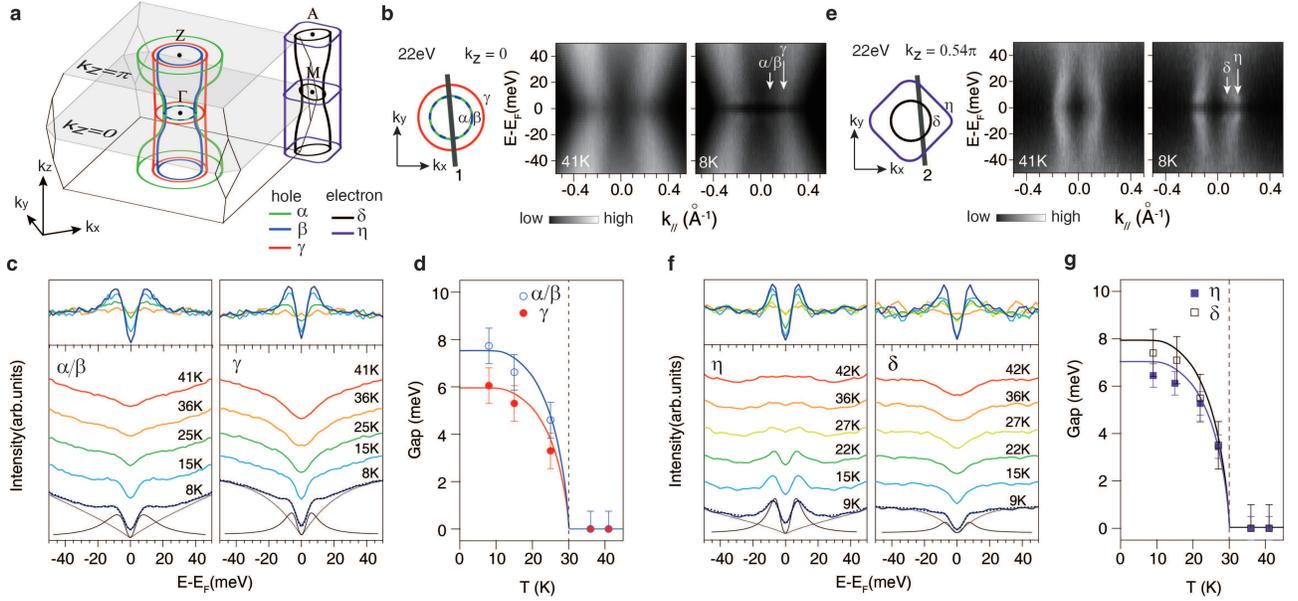}
\caption{\textbf{The temperature dependence of the photoemission
spectra of BaFe$_2$(As$_{0.7}$P$_{0.3}$)$_2$}. \textbf{a},  The
three-dimensional Fermi surfaces  of
BaFe$_2$(As$_{0.7}$P$_{0.3}$)$_2$. The two-iron unit cell are
implemented here, with Fe-Fe direction as the $k_x$ direction. The
electron Fermi surfaces are only illustrated at one corner of the
Brillouin zone for simplicity. \textbf{b},  The photoemission
intensity taken around the zone center with 22~eV photons at 41 and
8~K. \textbf{c}, The temperature dependence of the symmetrized
spectra measured at the Fermi crossings of $\alpha/\beta$ and
$\gamma$ along the cut in \textbf{b}. The top panel are  the low
temperature symmetrized spectra after subtracting the 41~K one.
\textbf{d}, The temperature dependence of the superconducting gaps
of the $\alpha/\beta$ and $\gamma$ band. The gap size is estimated
by fitting the symmetrized spectra as described in the supplementary
materials. Two examples of the fitted curve, background, and
superconducting spectral function are shown in black dashed line,
dot line, and solid line respectively. \textbf{e}, \textbf{f}, and
\textbf{g} are the same as \textbf{b}, \textbf{c}, and \textbf{d},
but taken around the zone corner along cut \#2 with 22~eV photons.
The error bars in \textbf{d} and \textbf{g} are standard deviations of the measured superconducting gaps.} \label{temperature}
\end{figure*}

The pairing symmetry of the Cooper pair in a superconductor is
manifested in its gap structure. Particularly, nodes or nodal lines
of the superconducting gap often imply unconventional (\textit{e.g.}
non-$s$-wave) pairing symmetries. For most iron-based
superconductors, there are  electron  Fermi surfaces at the
Brillouin zone corner and hole Fermi surfaces at the center. It has
been proposed that the  pairing interactions between the electron
and hole Fermi surfaces will induce nodeless $s$-wave order
parameter with  opposite signs on them \cite{Mazin,Kuroki,Hu0}.
While this nodeless  $s^{\pm}$-wave pairing symmetry  has gained
increasing  experimental  support
\cite{Tsueiphasesensitive,HanaguriSTM,Hu}, nodal gap has been
reported in  LaOFeP, LiFeP, KFe$_2$As$_2$,
BaFe$_2$(As$_{1-x}$P$_x$)$_2$, BaFe$_{2-x}$Ru$_x$As$_2$, and FeSe by
thermal conductivity, penetration depth, nuclear magnetic resonance,
and scanning tunneling spectroscopy studies \cite{LaFePO,linenode1,
linenode2,linenode3,LiFeP,ShiyanKFeAs,ShiyanBaRu,XueFeSeSTM}.
However,  no direct measurement on any of these compounds has been
reported regarding the gap structure so far, and especially the
location of the nodes remains unknown.  Since
BaFe$_2$(As$_{1-x}$P$_x$)$_2$ has relatively high superconducting
transition temperature $T_c$, it provides an opportunity for direct
access of the detailed gap structure in the momentum space by angle
resolved photoemission spectroscopy (ARPES).

We have conducted ARPES measurements on
BaFe$_2$(As$_{0.7}$P$_{0.3}$)$_2$ with a $T_c$ of 30~K (see Method
section for details). As previous detailed polarization dependent
studies have shown \cite{ZRYe} and replicated here in Fig.~1a, there
are three hole like Fermi surfaces ($\alpha$, $\beta$, and $\gamma$)
surrounding the central $\Gamma-Z$ axis of the Brillouin zone, and
two electron Fermi surfaces ($\delta$ and $\eta$) around the
Brillouin zone corner M-A. Near $\Gamma$, the $\alpha$ and $\beta$
Fermi surfaces are coincident, and these two bands are mainly
composed of Fe $d_{xz}$ and $d_{yz}$ orbitals near $\Gamma$. The
$\gamma$ band is composed of the $d_{xy}$ orbital, and shows little
$k_z$ dispersion \cite{ZRYe}. Compared with the Fermi surfaces of
Ba$_{0.6}$K$_{0.4}$Fe$_2$As$_2$ \cite{ZhangBK}, the Fermi surface of
the $\alpha$ band in BaFe$_2$(As$_{0.7}$P$_{0.3}$)$_2$ exhibits a
much larger warping away from $\Gamma$, so that it moves outward and
becomes the outmost Fermi surface around Z \cite{ZRYe,Fujimori}. To
examine the superconducting gap, Fig.~1b shows the symmetrized
photoemission intensity in the normal and superconducting state
taken along the cut \#1 across $\Gamma$. By symmetrizing, the
temperature broadening effect of the Fermi-Dirac distribution on the
spectrum can be minimized. By comparing the data at two different
temperatures, one could see that the spectral weight near the Fermi
energy ($E_F$) are suppressed in the superconducting state. More
specifically, the temperature dependence of the symmetrized
photoemission spectra at the Fermi crossings of the $\alpha$/$\beta$
and $\gamma$ bands are shown in Fig.~1c. In the normal state, there
is a spectral weight dip near the Fermi energy. It could be due to
the combined effect of fast dispersion nature of the band and
limited momentum resolution, or due to the intrinsic correlation
effects on the lineshape. Above $T_c$, such a dip does not change
noticeably from 41~K to 36~K. Upon entering the superconducting
state, the spectral weight near the Fermi energy is further
suppressed quickly with decreasing temperature. A small
superconducting coherent peak can be observed at 8~K. To remove the
effects of the normal state lineshape, the 41~K data are deducted
from the low temperature spectra, and it is clear that such a
suppression is related to the opening of a superconducting gap. By
fitting the spectra with a typical superconducting state spectral
function  (as exemplified in Fig.~1c and Fig.~S2 in the supplementary
materials), we could obtain the temperature dependence of the
superconducting gap in Fig.~1d, which can be fitted well to the BCS
formula (solid line). Similarly, the opening of the superconducting
gap on the $\delta$ and $\eta$ Fermi surfaces are illustrated in
Figs.~1e-1g for $k_z=0.54 \pi$. The extrapolated zero temperature
gaps give $2\Delta_0/k_B T_c \approx 5\sim6$ for
BaFe$_2$(As$_{0.7}$P$_{0.3}$)$_2$ . The $k_z$ is calculated and
folded into the upper half of the first Brillouin zone hereafter
(see supplementary materials for details). It is denoted in the unit
of $1/c'$ , where $c'$ is the distance between two neighboring FeAs
layers, which is half of the out-of-plane lattice constant $c$.

\begin{figure*}[t!]
\includegraphics[width=14cm]{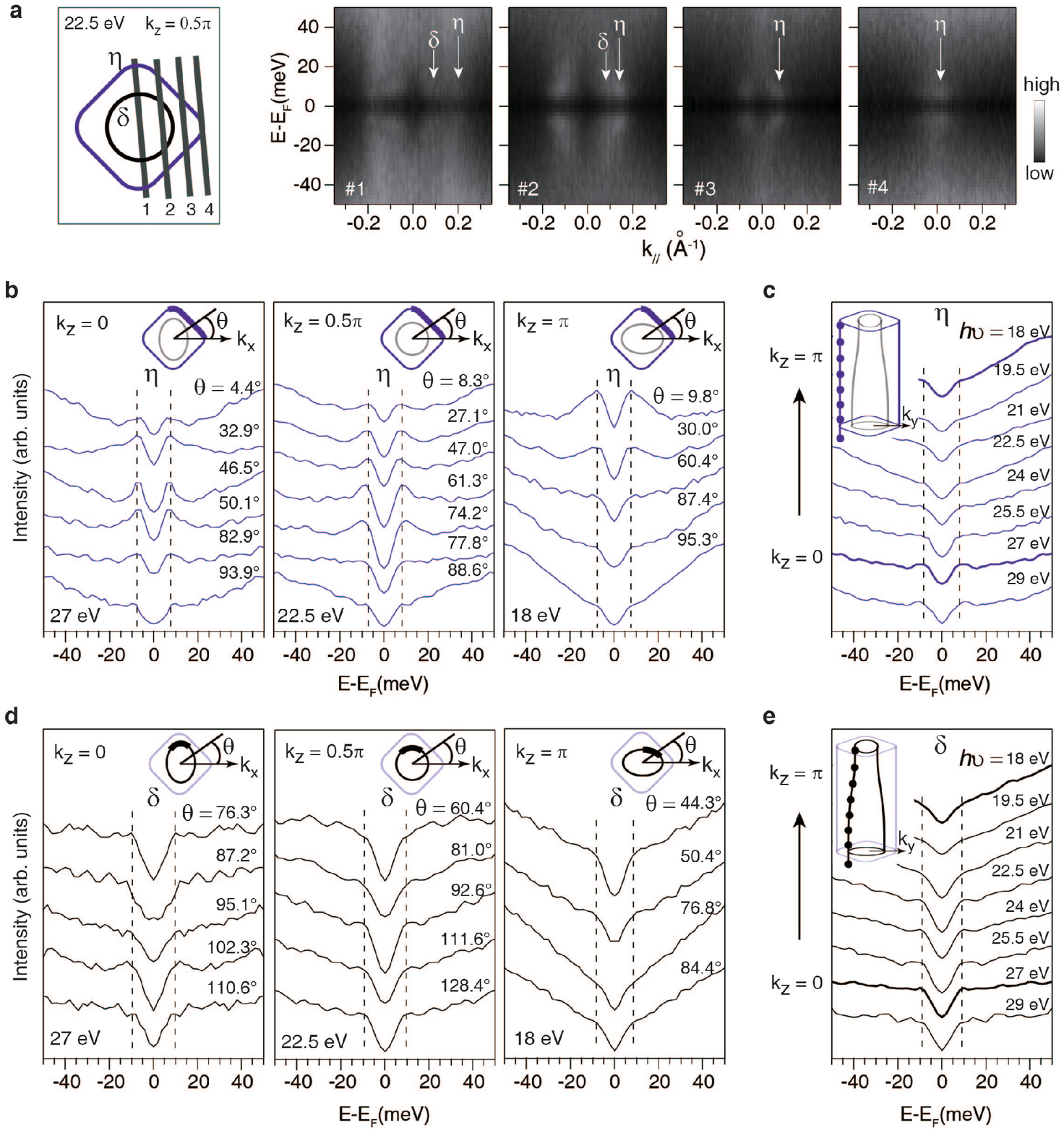}
\caption{\textbf{The superconducting gap distribution on the
electron Fermi surfaces of BaFe$_2$(As$_{0.7}$P$_{0.3}$)$_2$}.
\textbf{a},  The photoemission intensities taken along different
momentum cuts near the zone corner with 22.5~eV photons. \textbf{b},
Angular-dependent symmetrized spectra of the $\eta$ electron Fermi
surface measured at three typical $k_z$'s with 27, 22.5 and 18~eV
photons. \textbf{c}, $k_z$-dependence of the symmetrized spectra
measured on the $\eta$ electron Fermi surface. \textbf{d} and
\textbf{e} are the same as \textbf{b} and \textbf{c}, but taken on
the $\delta$ electron Fermi surface. The dashed line in \textbf{b},
\textbf{c}, \textbf{d}, \textbf{e} is a guide to the eyes for the
variation of the superconducting gap. Hereafter, the momentum
location of each spectrum  is denoted by either the polar angle
$\theta$ as defined in the insets, or a solid circle on the Fermi
surface. All data were taken in the superconducting state at 9~K.}
\label{M}
\end{figure*}

\begin{figure*}[t]
\includegraphics[width=16cm]{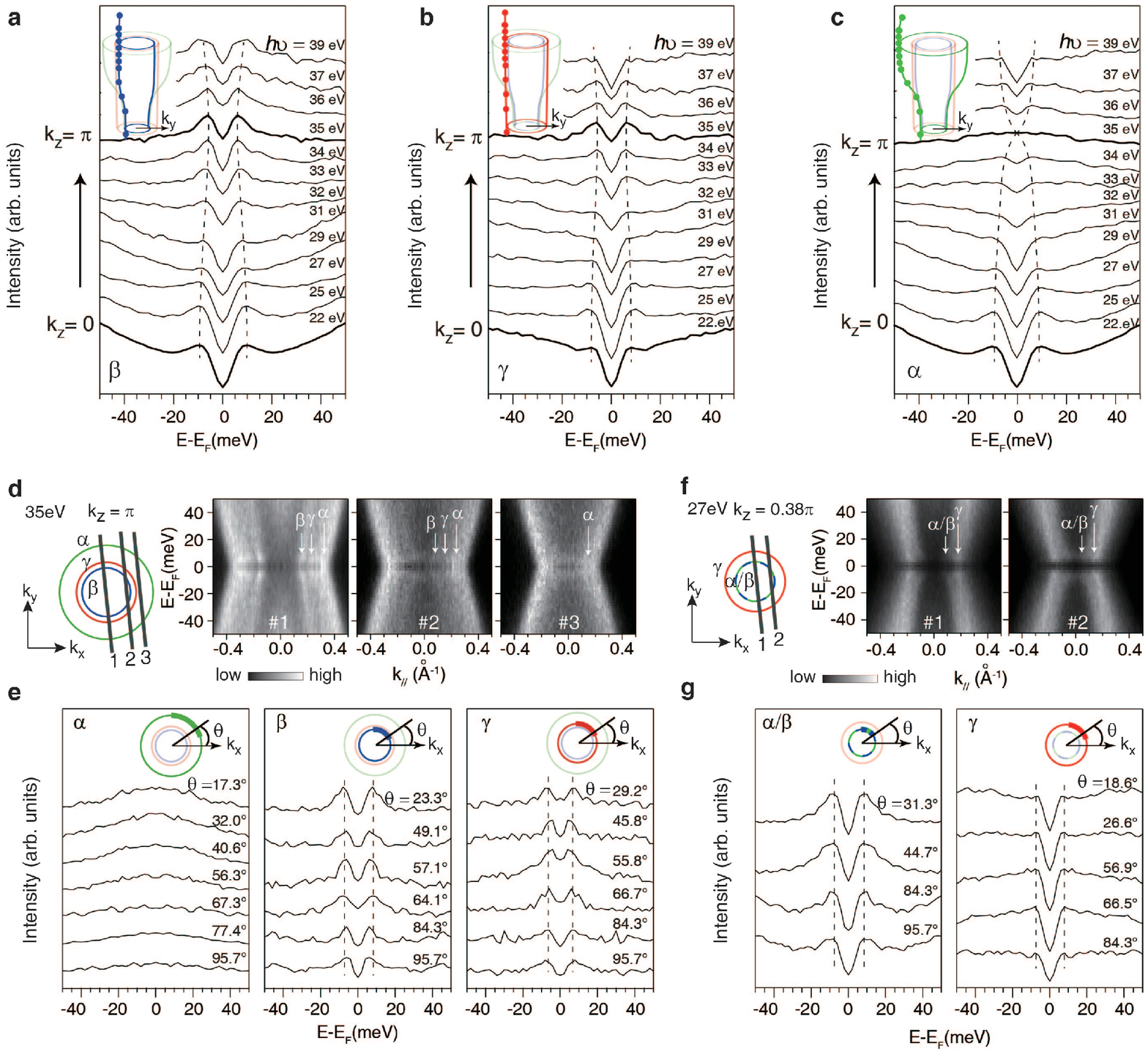}
\caption{\textbf{The superconducting gap distribution on the hole
Fermi surfaces of BaFe$_2$(As$_{0.7}$P$_{0.3}$)$_2$}.
\textbf{a}-\textbf{c}, $k_z$ dependence of the symmetrized spectra
measured on the $\beta$, $\gamma$, and $\alpha$ hole Fermi surfaces respectively. The symmetrized spectra near $k_z~=~0$ and $k_z~=~\pi$
are shown in thicker lines. The dashed line is a guide to the eyes
for  the variation of the superconducting gap at different $k_z$'s.
\textbf{d}, The photoemission intensities taken along three momentum
cuts and \textbf{e}, angular-dependent symmetrized spectra along the
$\alpha$, $\beta$, and $\gamma$ hole Fermi surfaces taken with 35~eV
photons in the $k_z=\pi$ plane around Z. \textbf{f} and \textbf{g}
are the same as \textbf{d} and \textbf{e}, but taken with 27~eV
photons in the $k_z=0.38\pi$ plane. All data were taken in the
superconducting state at 9~K. We note that one can usually
distinguish an energy gap if it is bigger than 20\% of the
resolution in a photoemission experiment, and the energy resolution
here with 35eV photons is 6 meV.} \label{Gamma}
\end{figure*}

Recent angle dependent transport experiments \cite{linenode2} have
suggested  that the superconducting gap nodes might be located on
the electron Fermi surfaces. Figure~2, together with more data in the supplementary materials, gives an extensive survey of gap on the
electron Fermi surfaces. Figure~2a shows four typical cuts in the
$k_z=0.5 \pi$ plane, and finite gaps clearly open in all cases. The
superconducting state spectra taken along the normal state  $\eta$
Fermi surface are shown in Fig.~2b for its $k_z=0$, $0.5 \pi$, and
$\pi$ horizontal cross-sections, and in Fig.~2c for its $k_z-k_y$
vertical cross-section. Similarly, Figs.~2d and 2e plot the spectra
on the $\delta$ Fermi surface.  In addition, similar data taken in
the $k_z=0.2\pi$, $0.34\pi$, $0.54\pi$, $0.64\pi$, $0.8\pi$ planes
are shown in Fig.~S4 of the supplementary materials. The
superconducting gaps are finite in all cases, and the in-plane gap
distribution is always isotropic. If the gap nodes were continuously
located along certain vertical lines or loops  on the Fermi surface
as suggested \cite{linenode2,dxy3,dxy4,dxy5}, they should not be
missed in such an extensive search. Therefore, we could conclude
that the line nodes of the gap are absent on the electron Fermi
surfaces.

Phosphor doping is predicted to alter the band structure and Fermi
surface topology dramatically \cite{Vildosola, dxy5}. Particularly,
it is predicted that the $d_{3z^2-r^2}$-based band would go above
$E_F$, while the $d_{xy}$-based band would move down below the $E_F$
with phosphor doping.  Several theories predicted that vertical line
nodes will appear in the electron Fermi surface when the $d_{xy}$
hole Fermi surface disappears \cite{dxy3, dxy4, dxy5}.  The fact
that the $d_{xy}$ band (\textit{i.e.} the $\gamma$ band) does not
sink below $E_F$,  and the absence of vertical nodes on the electron
Fermi surfaces all indicate that this is not the case in
BaFe$_2$(As$_{1-x}$P$_x$)$_2$.

\begin{figure*}[t]
\includegraphics[width=17cm]{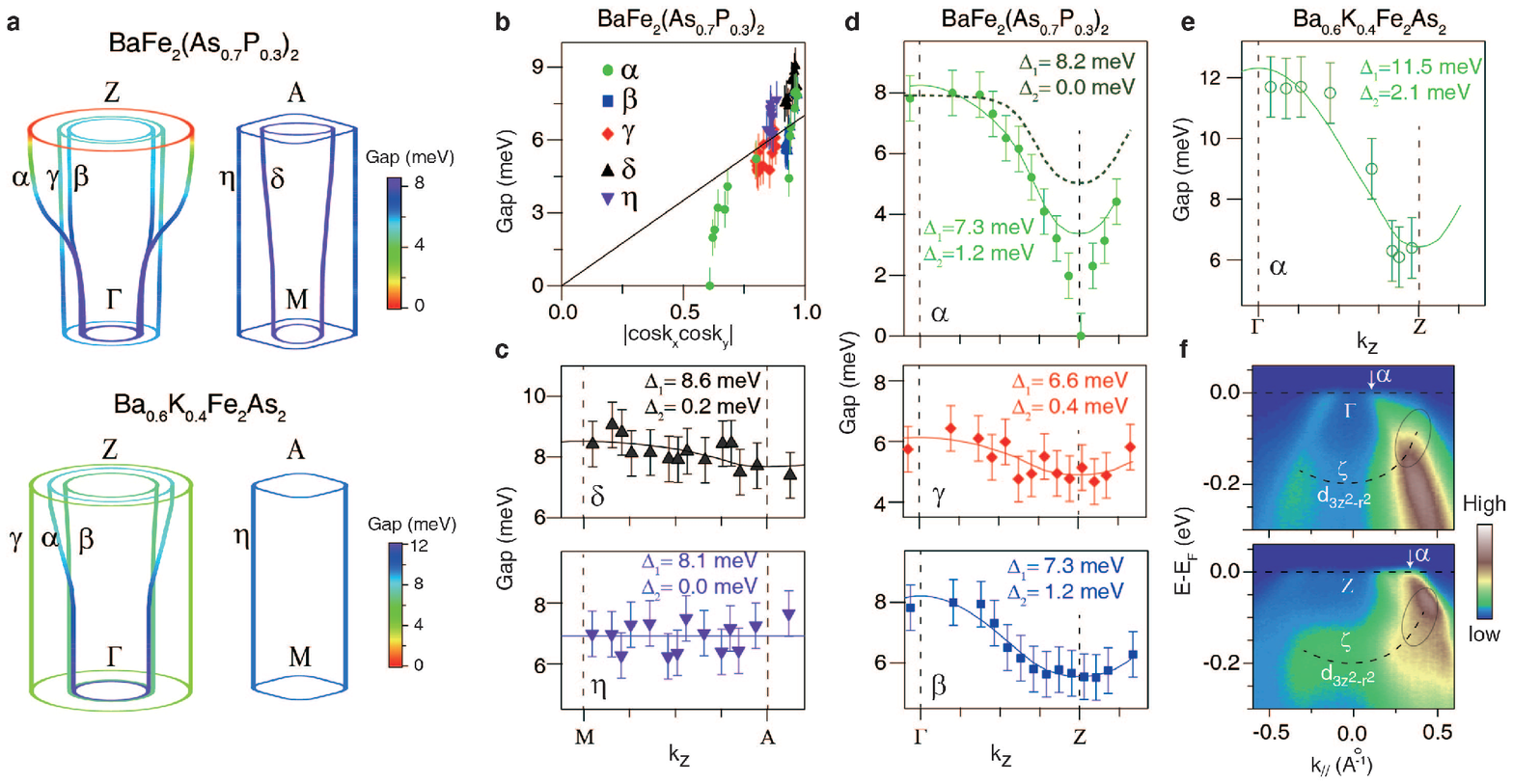}
\caption{\textbf{Momentum dependency of the superconducting gap.}
\textbf{a}, False color plots of the gap distribution on the Fermi
surfaces of BaFe$_{2}$(As$_{0.7}$P$_{0.3}$)$_2$ (top) and
Ba$_{0.6}$K$_{0.4}$Fe$_2$As$_2$ (bottom).  \textbf{b} The
superconducting gap of BaFe$_{2}$(As$_{0.7}$P$_{0.3}$)$_2$  vs. $|
\cos k_x \cos k_y |$. \textbf{c}, The superconducting gap on the
$\delta$ and  $\eta$ electron Fermi surfaces, and \textbf{d} on the
$\alpha$, $\beta$, and $\gamma$ hole Fermi surfaces  for
BaFe$_{2}$(As$_{0.7}$P$_{0.3}$)$_2$ as functions of $k_z$.
\textbf{e},  The superconducting gap on the $\alpha$ Fermi surface
of  Ba$_{0.6}$K$_{0.4}$Fe$_2$As$_2$ as a function of $k_z$. The error
bars are standard deviations of the measured superconducting gaps.
\textbf{f}, The band structure of BaFe$_{2}$(As$_{0.7}$P$_{0.3}$)$_2$ near $\Gamma$ and Z. }\label{sum}
\end{figure*}

% discuss the gap size etc

Figure~3 gives the detailed map of superconducting gaps on the hole
Fermi surfaces. The symmetrized spectra along the $\beta$, $\gamma$,
and $\alpha$ Fermi surfaces are shown in Figs.~3a, 3b and 3c
respectively  as a function of $k_z$. As indicated by the dashed
line,  the superconducting gap decreases slightly from $k_z=0$ to
$k_z= \pi$ for both the $\beta$ and $\gamma$ bands. Remarkably, the
gap of $\alpha$ exhibits a much larger variation than the other two,
and it even closes near the Z point. To further highlight the nodal
gap distribution, the photoemission intensity along three cuts in
the $k_z= \pi$  plane are shown in Fig.~3d, it is clear that  the
gap opens on the $\beta$ and $\gamma$ bands, but not on the $\alpha$
band in the data taken simultaneously. Figure~3e further displays
the symmetrized spectra around the three Fermi surfaces in the
$k_z=\pi$ plane. The gap exhibits a circular line node around the Z
point on the $\alpha$ Fermi surface, while the gaps on the $\beta$
and $\gamma$ Fermi surfaces are finite and isotropic within the
experimental resolution. As a comparison, the data taken in the
$k_z=0.38 \pi$ plane shows full gap for all three Fermi surfaces
(Figs.3f-3g). Such a horizontal line node distribution immediately
rules out the $d$-wave pairing symmetry, which would give four
vertical line nodes in the diagonal directions like in the cuprates.

We note that our finding naturally explains the four-fold symmetry
of the angle dependent thermal conductivity \cite{linenode2}, since
the Fermi velocity along the circular nodal line is four-fold
symmetric. Moreover, a recent ARPES measurement with 7~eV laser on
BaFe$_2$(As$_{0.65}$P$_{0.35}$)$_2$  has claimed a nodeless gap
around Z \cite{Shin}. However based on our extensive photon energy
dependence study, 7~eV photon actually measures the region with
$k_z=0.34\pi$,  which is far from Z  (see Fig.~S3 in supplementary
materials for detail). Therefore, gap nodes cannot be accessed by
7~eV photons.

The gap distribution on all the Fermi surfaces of
BaFe$_2$(As$_{0.7}$P$_{0.3}$)$_2$ is summarized in Fig.~4a, together
with that of Ba$_{0.6}$K$_{0.4}$Fe$_2$As$_2$ replicated from
Ref.\cite{ZhangBK} for comparison. The superconducting gaps on the
$\alpha$, $\beta$, $\gamma$,  and $\delta$ Fermi surfaces vary with
$k_z$ in BaFe$_2$(As$_{0.7}$P$_{0.3}$)$_2$, while only the gap on
$\alpha$  Fermi surface exhibits significant $k_z$ dependence in
Ba$_{0.6}$K$_{0.4}$Fe$_2$As$_2$.  Moreover, the Fermi surface potion
affected by the gap nodes in BaFe$_2$(As$_{0.7}$P$_{0.3}$)$_2$ is
limited, which explains its relative high $T_c$. Figure~4b plots the
superconducting gap of BaFe$_2$(As$_{0.7}$P$_{0.3}$)$_2$ as a
function of  $|\cos k_x \cos k_y|$. There are obvious deviations
from the predicted gap function of $\Delta_0|\cos k_x \cos k_y|$
based on the $s^{\pm}$-wave pairing symmetry
\cite{Mazin,Kuroki,Hu0,Hu}, which is particularly large for the
$\alpha$ band.

For quantitative analysis, the superconducting gap on each
individual Fermi surface of BaFe$_{2}$(As$_{0.7}$P$_{0.3}$)$_2$ is
plotted in Figs.~4c and 4d as functions of $k_z$. To account for the
$k_z$ dependence, they are fitted by a phenomenological function:
$\Delta_1|\cos k_x\cos k_y| + \Delta_2\cos k_z$. To achieve a
reasonable fit, different $\Delta_1$ and $\Delta_2$ have to be
exploited for each band although $\Delta_1$ varies just slightly,
indicating the orbital dependency of the gap. However, the gap
structure of the $\alpha$ band cannot be fitted well by such a
formula (solid line) probably due to strong mixing of another
orbital into this band as will be discussed below. As a reference,
we plot the $k_z$ dependence of the gap of the $\alpha$ band in
Ba$_{0.6}$K$_{0.4}$Fe$_2$As$_2$ and its fit in Fig.~4e, where the
$k_z$-contribution to the gap reduction is about 4.2~meV from the
$\Delta_2$ term.

Enforcing a $\Delta_1|\cos k_x\cos k_y|$ fit (dashed line) to the
superconducting gap of the $\alpha$ band of
BaFe$_2$(As$_{0.7}$P$_{0.3}$)$_2$, one could estimate the gap with only in-plane momentum contribution.
It is about 5~meV around Z (bottom of the dashed line), due to the
large radius of the $\alpha$ Fermi surface in the $k_z=\pi$ plane.
On the other hand, considering the stronger three dimensional
character of the $\alpha$ band in
BaFe$_2$(As$_{0.7}$P$_{0.3}$)$_2$, it is therefore reasonable to
postulate that the $k_z$-related reduction of the gap could be even larger than that in Ba$_{0.6}$K$_{0.4}$Fe$_2$As$_2$. As a result, it further reduce the gap by another 5~meV to zero, and create the
observed horizontal circular node. That is, the circular nodal gap
surrounding Z in BaFe$_2$(As$_{0.7}$P$_{0.3}$)$_2$ could be an
``accident" conspired by the relatively small $\Delta_1$, the large
hole Fermi surface radius around Z, and the sizable $k_z$-related gap
reduction. It is possible that with increasing $k_z$-dependency, the
circular node could be reached at a smaller $k_z$, and the
superconducting order parameter would change sign beyond that $k_z$.

Our findings are consistent with a scenario proposed by Kuroki et
al. \cite{dxz}.  They found that with increasing  phosphor
concentration, 3$d_{3z^2-r^2}$ orbital will be strongly mixed into
the $\alpha$ band near Z. As a result, the $s^{\pm}$ superconducting
gap could change sign on the $\alpha$ Fermi surface near Z, and thus
a nodal circle of superconducting gap emerges on the location where
the sign is switched. Furthermore,  more phosphor doping will shift
the  nodal circle from Z toward $\Gamma$. Indeed as illustrated in
Fig.~4f and Fig.~S6 in the supplementary materials, we do find that the $\zeta$ band with strong
$d_{3z^2-r^2}$ orbital character comes cross the $\alpha$ band, and
the Fermi crossing of $\alpha$ is much closer to the top of the $\zeta$ band near Z
than near $\Gamma$. This would induce significant amount of
$d_{3z^2-r^2}$ orbital character into $\alpha$ near $E_F$.
This scenario together with our findings provides an possible
explanation for the recent intriguing observation that the nodal gap
will appear in iron pnictides when the distance between the
pnictogen and Fe plane ($h_{Pn}$) is smaller than 1.33\AA
\cite{LiFeP}. In the case of BaFe$_2$(As$_{1-x}$P$_x$)$_2$, $h_{Pn}$
is reduced by  phosphor doping, and consequently, it would cause larger
$k_z$-dispersion and strong mixing of the $d_{3z^2-r^2}$ orbital for the $\alpha$ band as observed, which would create the line node as proposed \cite{dxz}.

To summarize, we have mapped out the detailed superconducting gap
distribution of BaFe$_2$(As$_{0.7}$P$_{0.3}$)$_2$, and found that
the line node of superconducting gap is a circle around Z on the
$\alpha$ Fermi surface, which is likely induced by the strong three
dimensional nature of the $\alpha$ band such as its increasing
$d_{3z^2-r^2}$ orbital character near Z. Our results rule out the
$d$-wave pairing symmetry as the cause of the nodal
superconductivity, and indicate that $s^{\pm}$-wave pairing symmetry
still prevails here.  This unifies the seemingly diversified
phenomenology of nodal and nodeless superconducting gaps in various
iron based superconductors, and provides a discriminator for
theories on iron pnictides.

\textbf{Methods:} High quality
BaFe$_{2.1}$(As$_{0.7}$P$_{0.3}$)$_{1.9}$ single crystals with
superconducting transition temperature $T_c$~=~30~K  were
synthesized through flux-free crucible growth. Shiny platelet
crystals as large as 2$\times$ 2$\times$ 0.05~mm$^3$ were obtained
with residual resistivity ratio of about 10 (see supplementary
information for details). The compositions were measured by an
energy dispersive X-ray (EDX) analysis.  Data were taken at the
Beamline 5-4 of Stanford Synchrotron Radiation Lightsource (SSRL).
All the data were taken with Scienta electron analyzers, the overall
energy resolution was 5-8~meV depending on the photon energy, and
the angular resolution was 0.3 degree. The samples were cleaved
\textit{in situ}, and measured under ultra-high-vacuum of
5$\times$10$^{-11}$\textit{torr}.

\textbf{Acknowledgement:} The authors thank Prof. J. P. Hu, and
Prof. X. H. Chen for fruitful discussions, and thank Dr. D.H. Lu for the
experimental assistance at SSRL. This work is supported in part by the
National Science Foundation of China, Ministry of Education of
China, Science and Technology Committee of Shanghai Municipal, and
National Basic Research Program of China (973 Program)  under the
grant Nos. 2011CB921802 and 2011CBA00112. SSRL is operated by the US
DOE, Office of Basic Energy Science, Divisions of Chemical Sciences
and Material Sciences.

\end{document}

\title{Supplementary information for:

Nodal superconducting gap structure in ferropnictide superconductor  BaFe$_2$(As$_{0.7}$P$_{0.3}$)$_2$}

\author{Y. Zhang}\author{Z. R. Ye}\author{Q. Q. Ge}\author{F. Chen}\author{Juan Jiang}\author{M. Xu}\author{B. P. Xie}\author{D. L. Feng}\email{dlfeng@fudan.edu.cn}
\affiliation{State Key Laboratory of Surface Physics, Department of
Physics,  and Advanced Materials Laboratory, Fudan University,
Shanghai 200433, People's Republic of China}

\maketitle

\section{Sample Description}

High quality BaFe$_{2}$(As$_{1-x}$P$_x$)$_2$ ($x$~=~0.30) single
crystals were synthesized with a flux-free method. Ba, FeAs and FeP
were mixed with the nominal compositions, loaded into an alumina
tube, and then sealed into a stainless steel crucible under the Ar
atmosphere. The entire assembly was heated to 1673~K  and kept for
12~h or longer, and then slowly cooled  down to 1173~K at the rate
of 4~K/h before shutting off the power. Shining platelet crystals as
large as 2$\times$ 2$\times$ 0.05~mm$^3$ were obtained as shown in
the inset of Fig.~S1a. The resistivity of
BaFe$_2$(As$_{0.7}$P$_{0.3}$)$_2$ single crystal shows a sharp
superconducting transition at 30~K, with the transition width less
than 0.5~K (Fig.~S1a). The room temperature-to-residual resistivity
ratio is $\approx$ 10.4, indicating minimal disorder and impurity of
the single crystal. The magnetic susceptibility as a function of
temperature was measured in a 20~Oe magnetic field (Fig.~S1b). A
sharp drop in the zero field cooling (ZFC) susceptibility,
indicating the onset of superconductivity, appears at 30K,
which is consistent with the zero-resistivity temperature in
Fig.~S1a. The ZFC and field cooling (FC) curves indicate a bulk
superconductivity nature and high quality of the crystals.

\begin{figure*}[h]
\includegraphics[width=15cm]{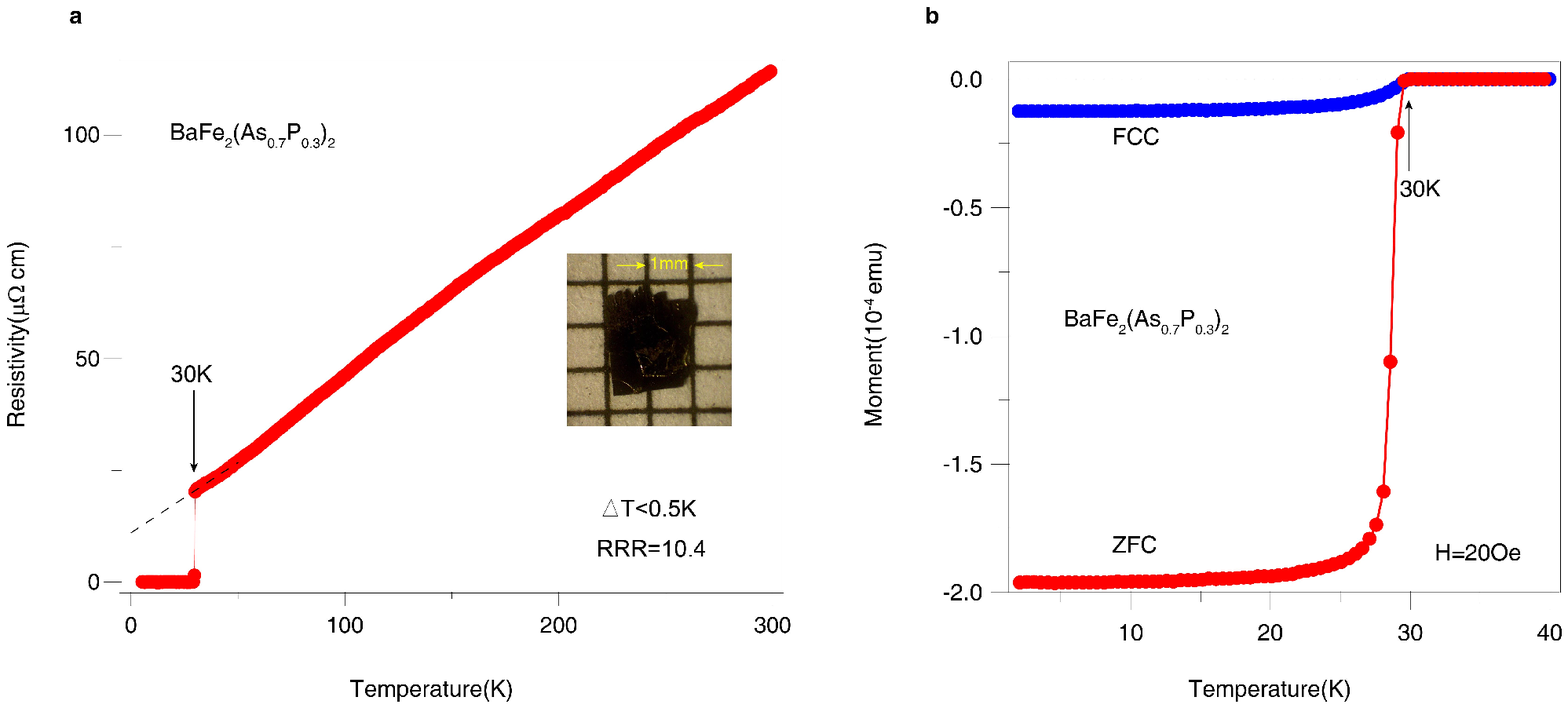}
\caption{\textbf{Resistivity and magnetic susceptibility
 of BaFe$_2$(As$_{0.7}$P$_{0.3}$)$_2$}. \textbf{a}, Temperature dependence of  resistivity.
The inset is the photograph of a
BaFe$_2$(As$_{0.7}$P$_{0.3}$)$_2$ single crystal. \textbf{b},
Temperature dependence of magnetic susceptibility in a 20~Oe
magnetic field.} \label{trans}
\end{figure*}

\section{Determination of the superconducting gap}

To determine the gap size, we fit the symmetrized spectra or energy distribution curves (EDCs) with a simple empirical spectral function described in Ref.~1. The self-energy at the normal state Fermi crossing ($k_F$) in the superconducting state may be expressed as

\[\Sigma ({k_F},\omega ) =  - i\Gamma  + \frac{{{\Delta ^2}}}{\omega }\]

where $\Gamma$ is the inverse life time of the quasi-particle related to the width of the spectrum and $\Delta$
represents the gap at $k_F$. The spectral function is calculated according to

\[A({k_F},\omega ) =  - \frac{1}{\pi }{\mathop{\rm Im}\nolimits} \frac{1}{{\omega  - \Sigma ({k_F},\omega )}}\]

The spectral function  was convoluted with  a Gaussian representing the instrumental resolution. The spectral background is fitted by a fourth order polynomial. We note that some variations of the smooth background  has minimal effect on the fitted superconducting gap size. The gap is most determined by the peak position near the Fermi energy ($E_F$) in the symmetrized EDCs. Due to the broad line shape and small spectral weight of the superconducting peak here, the  superconducting gap determined through the fitting procedure is usually about 1~meV smaller than the gap value determined directly from the superconducting peak position. As an example, the fit of the data in the left panel of Fig.~1c are shown in Fig.~S2, which gives the gap value of the $\alpha/\beta$ band as a function of temperature.

\begin{figure*}[h]
\includegraphics[width=14cm]{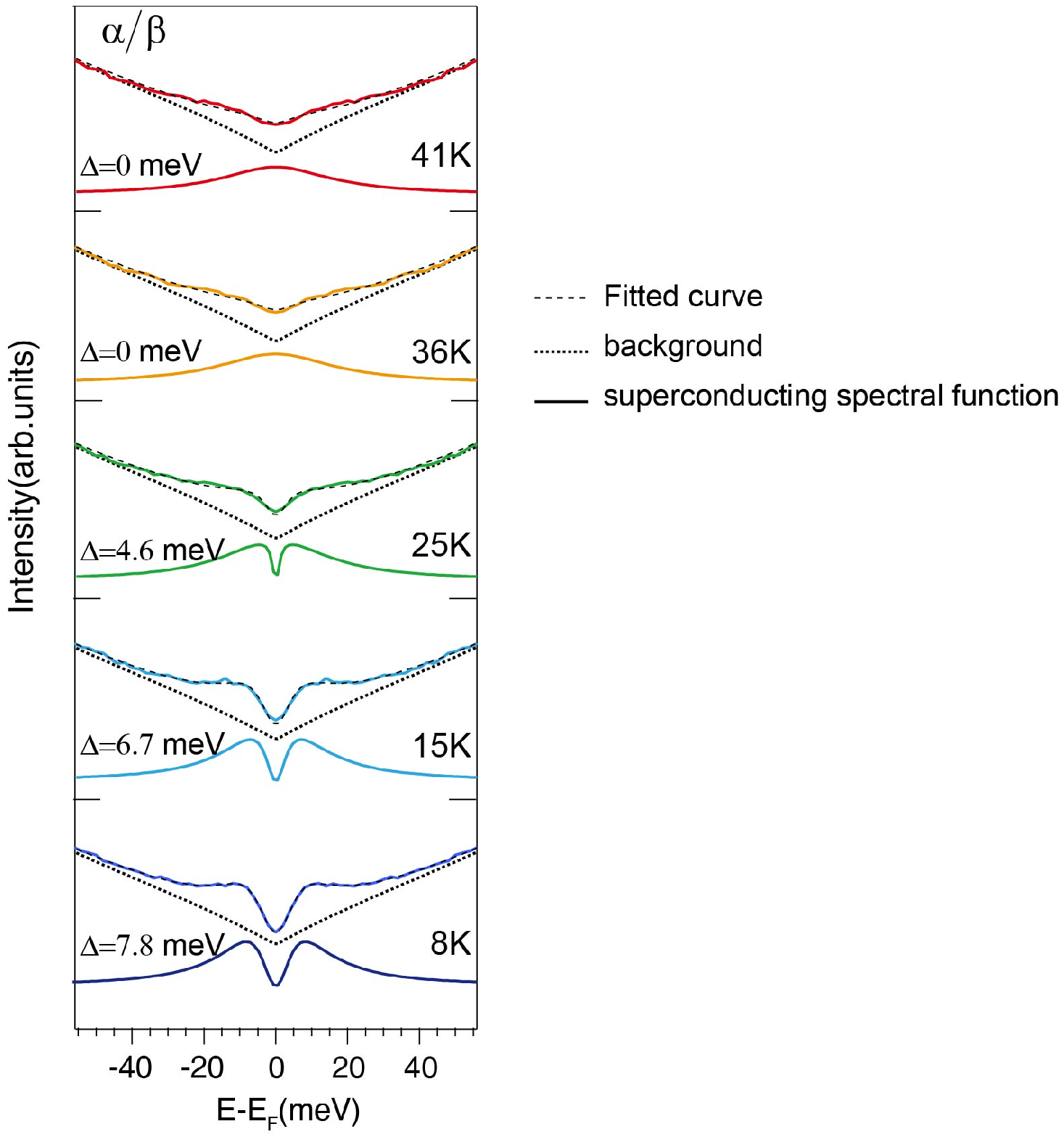}
\caption{\textbf{An empirical fit for retrieving the superconducting gap value of BaFe$_2$(As$_{0.7}$P$_{0.3}$)$_2$.}  The symmetrized spectra measured at different temperatures
are fitted by the empirical spectral function. Data were taken at a Fermi crossing  of the $\alpha/\beta$ Fermi surface near $\Gamma$ (also shown in Fig.~3c in the main text).
The fitted curve, background, and superconducting spectral function are shown in black dashed line, dot line, and solid line respectively.} \label{fit}
\end{figure*}

\newpage
\section{Determination of $k_z$}

In angle resolved photoemission spectroscopy, a specific out-of-plane momentum $k_z$ can be accessed with a particular photon energy according to

\[{k_z} = \sqrt {\frac{{2m}}{{{\hbar ^2}}}[(h\nu  - \Phi ){{\cos }^2}\theta  + {V_0}]} \]

where $h\nu$ is the photon energy, $\Phi$ is the work function, and $V_0$ is the inner potential of the sample \cite{hufner}. To determine $V_0$, we plot the photon energy dependence of the momentum distribution curves (MDCs) near $E_F$  along high symmetry directions in the Brillouin zone in Fig.~S3. The position of the MDCs represents the band dispersion as a function of $k_z$, and one could search for the symmetry in the band dispersion enforced by the Brillouin boundary. For example in Fig.~S3a, the peak position of $\alpha$ is outmost at 35~eV, representing the largest Fermi surface at the Z point. Through trial and error, we finally pick the inner potential of 14.5~eV to calculate the $k_z$ in order to achieve best agreement with all our photon energy dependent data in Fig.~S3. The determination of $k_z$'s for different photon energies are shown in Fig.~S3c, where the cuts taken with 22, 35, 27, and 18~eV photons sample momentum regions near $\Gamma$, Z, M, and A, respectively. Based on our $V_0$, the $k_z$ is about 0.34$\pi$ for the cut taken with 7~eV photons.
Note that, the $k_z$ is folded into the upper half of the first Brillouin zone for simplicity in the main text.

\begin{figure*}[h]
\includegraphics[width=16cm]{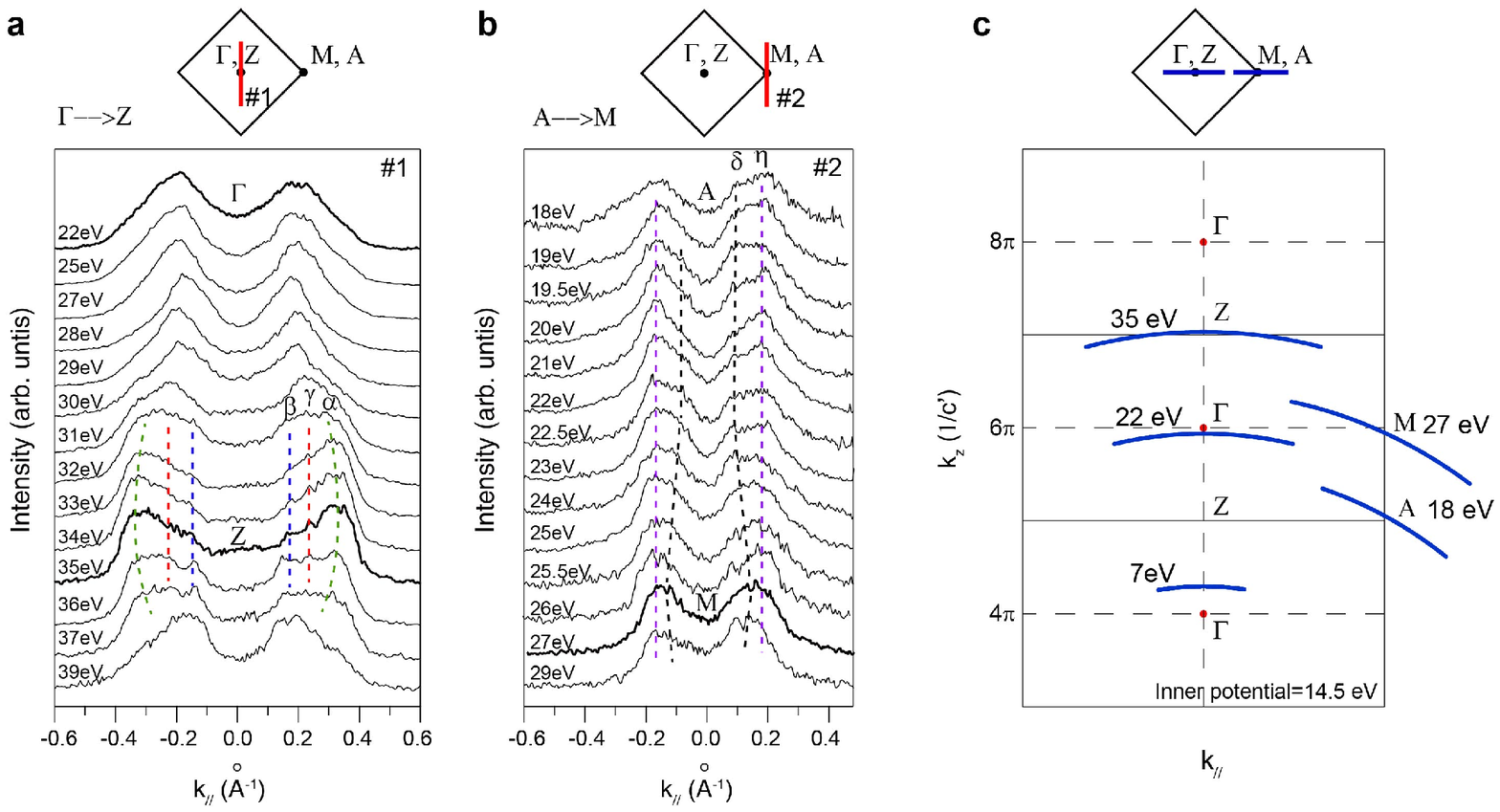}
\caption{\textbf{Photon energy dependent data of BaFe$_2$(As$_{0.7}$P$_{0.3}$)$_2$.} \textbf{a}, The photon energy dependent momentum distribution curves (MDCs) near $E_F$ taken along cut \#1. The peak positions of the MDCs represent the Fermi crossings of the hole Fermi surfaces near the zone center.  \textbf{b} is the same as \textbf{a} but taken along cut\#2, where the peak positions of the MDCs represent the Fermi crossings of the electron Fermi surfaces near the zone corner. \textbf{c}, The illustration of the $k_z$ calculation for the cuts taken with 22, 35, 27, 18, and 7~eV photons.} \label{calkz}
\end{figure*}

\newpage

\section{Symmetrized EDCs on the electron and hole Fermi surfaces}

In addition to the data presented in Fig.2,  Fig.~S4 present more superconducting gap measurement on the electron Fermi surfaces of BaFe$_2$(As$_{0.7}$P$_{0.3}$)$_2$ in the $k_z=0.2\pi$, $0.34\pi$, $0.54\pi$, $0.64\pi$, $0.8\pi$ planes. The superconducting gap is finite in every symmetrized EDC. The extensive survey gives compelling evidence on the absence of line node on the two electron Fermi surfaces in BaFe$_2$(As$_{0.7}$P$_{0.3}$)$_2$.

\begin{figure*}[h]
\includegraphics[width=17cm]{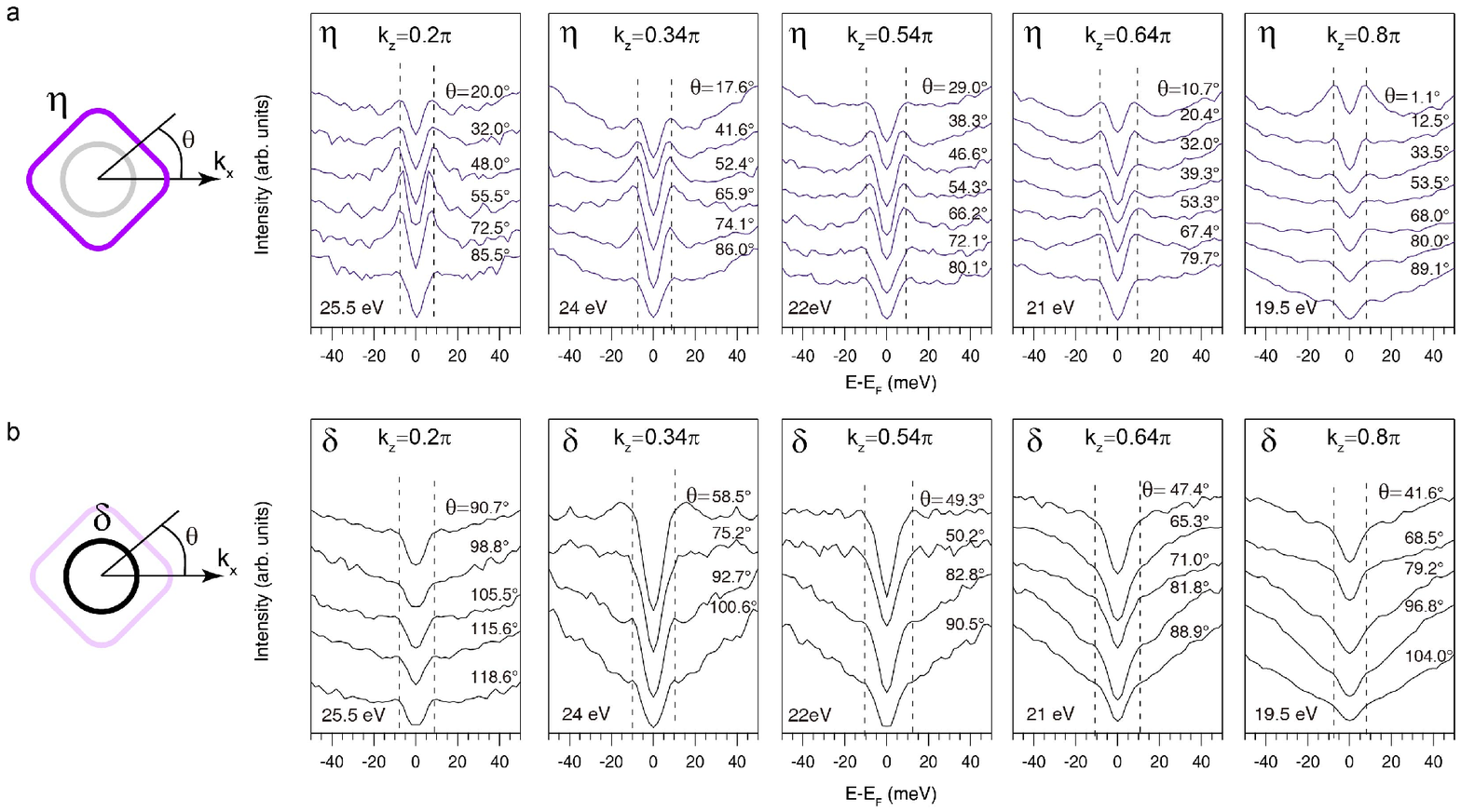}
\caption{\textbf{The symmetrized EDCs on the electron Fermi surfaces.}
\textbf{a}, The angular-dependent symmetrized EDCs measured on the $\eta$ electron Fermi surfaces with different photon energies.  \textbf{b}, The angular-dependent symmetrized EDCs measured on the $\delta$ electron Fermi surfaces with different photon energies.} \label{M}
\end{figure*}

In addition to the data presented in Fig.~3g, Fig.~S5 present more superconducting gap measurement on the hole Fermi surfaces of BaFe$_2$(As$_{0.7}$P$_{0.3}$)$_2$ in the $k_z=0$ plane. The gap distribution is isotropic and finite here.

\begin{figure*}[h]
\includegraphics[width=7cm]{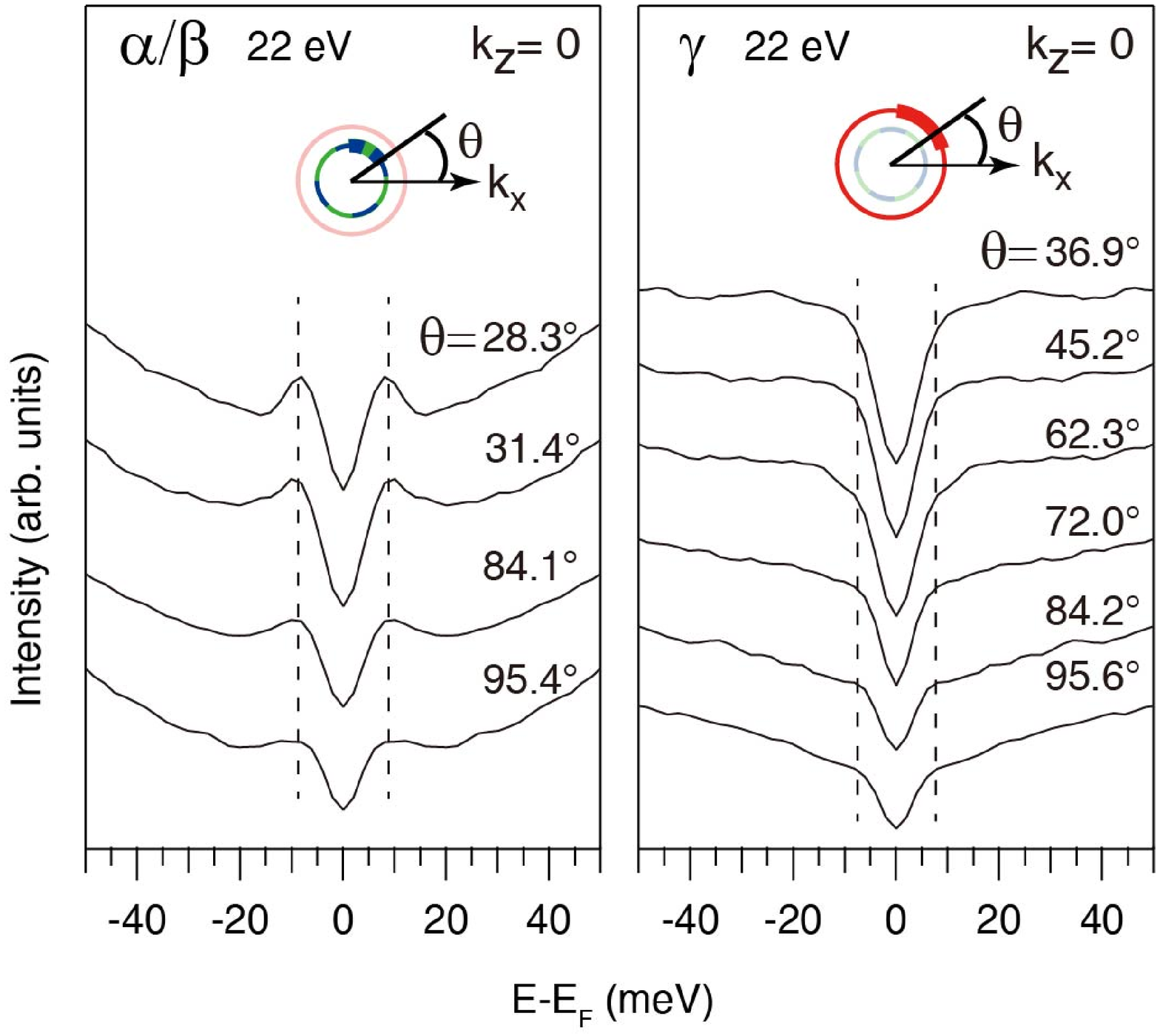}
\caption{\textbf{The symmetrized EDCs on the hole Fermi surfaces.} The angular-dependent symmetrized EDCs measured on the hole Fermi surfaces with 22~eV photons near the $\Gamma$ point.} \label{M}
\end{figure*}

\newpage

\section{Stronger mixing of $d_{z^2}$ orbital into the $\alpha$ band near Z}

Previous experimental and theoretical studies on the electronic structure of iron-based superconductors report that the $d_{3z^2-r^2}$ orbital form two bands that are below $E_F$ \cite{Zhangorbital, bandthoery}. In BaFe$_2$(As$_{0.7}$P$_{0.3}$)$_2$, the $\zeta$ band with $d_{3z^2-r^2}$ orbital is at about 200~meV below $E_F$ near the $\Gamma$ point as shown in Figs.~S6a. Near $\Gamma$, one could trace the MDCs peaks correspond to $\zeta$ disperse vertically  towards $E_F$ as marked by the solid squares in Fig.~S6b at about $k_{\parallel}~\approx~0.3~\AA^{-1}$. This indicates these are just the residual of the broad $\zeta$ feature below $E_F$. In other words, although the $\zeta$ band does not cross $E_F$, it could contribute finite spectral weight even at $E_F$. The $\alpha$ band crosses $E_F$ at 0.1~$\AA^{-1}$ near $\Gamma$, and at 0.33~$\AA^{-1}$ near Z (Figs.~S6a and S6c). In the MDCs taken around Z shown in Fig.~S6d, the features of $\alpha$ and those of $\zeta$ are too close to be distinguished. Thus, the $d_{3z^2-r^2}$ orbital character in the $\zeta$ band should mix rather strongly with the $\alpha$ band near $E_F$ around the Z point \cite{zirong}.

\begin{figure*}[h]
\includegraphics[width=10cm]{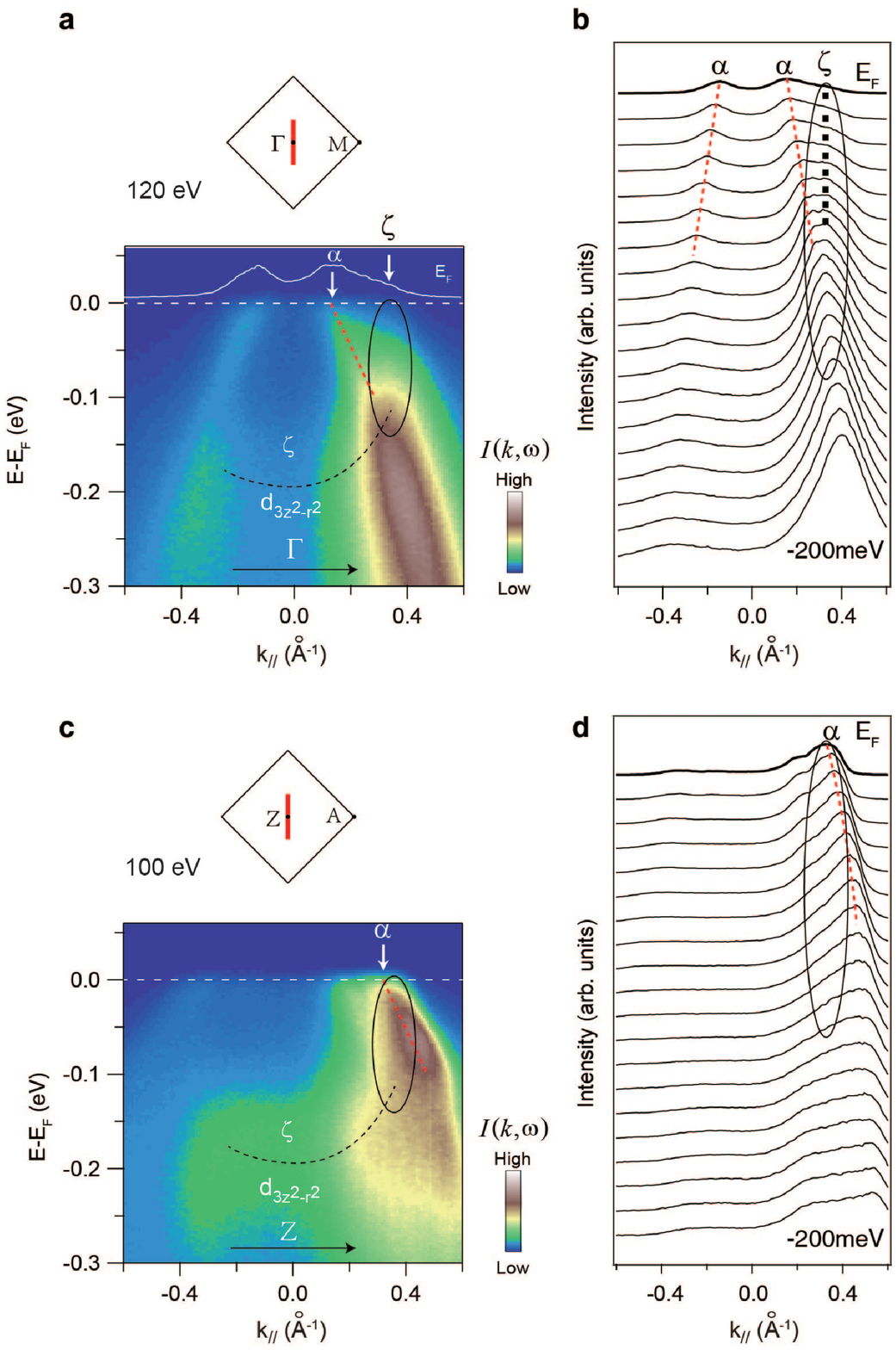}
\caption{\textbf{The photoemission data of BaFe$_2$(As$_{0.7}$P$_{0.3}$)$_2$ near $E_F$ taken in
the $p$ polarization \cite{zirong}.} \textbf{a}, The photoemission
intensity  taken with 120~eV
photons around $\Gamma$ in the $p$ polarization. Dashed
lines are band dispersions. The black ellipse indicates the spectral weight of the $\zeta$ band. The inset in
\textbf{a} is the MDC at $E_F$. \textbf{b},  The   MDCs of the data in \textbf{a}. \textbf{c}, The photoemission
intensity  taken with 100~eV photons around Z in the $p$ polarization.  \textbf{d} The  MDCs of the data in \textbf{c}.} \label{dz2}
\end{figure*}